\documentstyle[epsfig,12pt,a4p]{article}

\newcommand{\chic}{\mbox{$\chi_c$ }}

\newcommand{\etapipi}{\mbox{$\eta \pi^+\pi^-$ }}

\newcommand{\kkpi}{\mbox{$K^0_SK^\pm \pi^\mp$ }}

\begin{document}
\begin{titlepage}
\def\footnoterule{\hrule width 1.0\columnwidth}
%\hfill  \hfill
%6\thinspace February\thinspace 1990
% \begin{center} {\large EUROPEAN ORGANIZATION FOR NUCLEAR RESEARCH}
%  \end{center}
\begin{tabbing}
put this on the right hand corner using tabbing so it looks
 and neat and in \= \kill
\> {6 June 2000}
\end{tabbing}
\bigskip
\bigskip
\begin{center}{\Large  {\bf A search for charmonium states
produced in central
pp interactions at 450 GeV/c}
}\end{center}

\bigskip
\bigskip
\begin{center}{        The WA102 Collaboration
}\end{center}\bigskip
\begin{center}{
D.\thinspace Barberis$^{  4}$,
F.G.\thinspace Binon$^{   6}$,
F.E.\thinspace Close$^{  3,4}$,
K.M.\thinspace Danielsen$^{ 11}$,
S.V.\thinspace Donskov$^{  5}$,
B.C.\thinspace Earl$^{  3}$,
D.\thinspace Evans$^{  3}$,
B.R.\thinspace French$^{  4}$,
T.\thinspace Hino$^{ 12}$,
S.\thinspace Inaba$^{   8}$,
A.\thinspace Jacholkowski$^{   4}$,
T.\thinspace Jacobsen$^{  11}$,
G.V.\thinspace Khaustov$^{  5}$,
J.B.\thinspace Kinson$^{   3}$,
A.\thinspace Kirk$^{   3}$,
A.A.\thinspace Kondashov$^{  5}$,
A.A.\thinspace Lednev$^{  5}$,
V.\thinspace Lenti$^{  4}$,
I.\thinspace Minashvili$^{   7}$,
J.P.\thinspace Peigneux$^{  1}$,
V.\thinspace Romanovsky$^{   7}$,
N.\thinspace Russakovich$^{   7}$,
A.\thinspace Semenov$^{   7}$,
P.M.\thinspace Shagin$^{  5}$,
H.\thinspace Shimizu$^{ 10}$,
A.V.\thinspace Singovsky$^{ 1,5}$,
A.\thinspace Sobol$^{   5}$,
M.\thinspace Stassinaki$^{   2}$,
J.P.\thinspace Stroot$^{  6}$,
K.\thinspace Takamatsu$^{ 9}$,
T.\thinspace Tsuru$^{   8}$,
O.\thinspace Villalobos Baillie$^{   3}$,
M.F.\thinspace Votruba$^{   3}$,
Y.\thinspace Yasu$^{   8}$.
%% \end authorlist
}\end{center}

\begin{center}{\bf {{\bf Abstract}}}\end{center}

{
A search for centrally produced charmonium states has been presented.
There
is no significant
evidence for any charmonium production.
An upper limit of 2~nb is found for the cross section of
\chic production using the decay
$\chi_c(1P)\rightarrow J/\psi\gamma$.
}
\bigskip
\bigskip
\bigskip
\bigskip\begin{center}{{Submitted to Physics Letters}}
\end{center}
%\newpage
\bigskip
\bigskip
\begin{tabbing}
aba \=   \kill
% $^\dag$ \> \small
% Deceased. \\
$^1$ \> \small
LAPP-IN2P3, Annecy, France. \\
$^2$ \> \small
Athens University, Physics Department, Athens, Greece. \\
%% $^3$ \> \small
%% Bergen University, Bergen, Norway. \\
$^3$ \> \small
School of Physics and Astronomy, University of Birmingham, Birmingham, U.K. \\
$^4$ \> \small
CERN - European Organization for Nuclear Research, Geneva, Switzerland. \\
$^5$ \> \small
IHEP, Protvino, Russia. \\
$^6$ \> \small
IISN, Belgium. \\
$^7$ \> \small
JINR, Dubna, Russia. \\
$^8$ \> \small
High Energy Accelerator Research Organization (KEK), Tsukuba, Ibaraki 305-0801,
Japan. \\
$^{9}$ \> \small
Faculty of Engineering, Miyazaki University, Miyazaki 889-2192, Japan. \\
$^{10}$ \> \small
RCNP, Osaka University, Ibaraki, Osaka 567-0047, Japan. \\
$^{11}$ \> \small
Oslo University, Oslo, Norway. \\
$^{12}$ \> \small
Faculty of Science, Tohoku University, Aoba-ku, Sendai 980-8577, Japan. \\
\end{tabbing}
\end{titlepage}
\setcounter{page}{2}
\bigskip
\par
The WA102 collaboration
have studied centrally produced
final states formed  in the reaction
\begin{equation}
pp \rightarrow p_{f} (X^0) p_{s}
\label{eq:a}
\end{equation}
at 450 GeV/c.
The subscripts $f$ and $s$ indicate the
fastest and slowest particles in the laboratory respectively.
By measuring the cross section as a function of energy
it has been possible to deduce that a large
number of the final states are compatible with being produced by
Double Pomeron Exchange (DPE).
Apart from kinematical factors DPE should be effectively
flavour blind in the production of resonance states.
However, Donnachie and Landshoff~\cite{DL} have recently
claimed that in order
to describe data from HERA they need to introduce two Pomerons; a so-called
soft Pomeron with y axis intercept at 1.08
on the Chew-Frautschi plot and a hard Pomeron with intercept
at 1.4. In addition, they have claimed~\cite{DL2} that the soft Pomeron
has a very weak coupling to $c \overline c$ pairs.
\par
To date no evidence has been observed for
charmonium production
in DPE.
These states will be heavily suppressed due to
the mass reach of the experiment,
%which is given by $M^2$~=~$sx_1x_2$, where
%$M$ is the mass of the central system, $s$ is the centre of mass energy
%squared and $x_1,x_2$ are the Feynman x of the Pomerons
%which are typically in the range 0. to 0.1.
%Hence at a centre of mass energy of 29.1~GeV the available phase space
%for the production of resonances is running out by 3.0~GeV.
hence a search for charmonium states is limited to the lightest.
Possible candidates are the $\eta_c(1S)$, the $J/\psi$ and the
$\chi_c(1P)$ states.
\par
The $J/\psi$ can not be exclusively produced in DPE due to C-parity.
It has been observed previously that
$J^{PC}$~=~$0^{-+}$ states are suppressed in DPE~\cite{0mpap}
therefore it is likely that
the $\eta_c(1S)$
is also suppressed.
$J^{PC}$~=~$0^{++}$, $1^{++}$ and $2^{++}$ states are seen prominently in DPE
therefore it may be interesting to search for the \chic states.
The dominant decay mode for the $\chi_1(3510)$ and $\chi_2(3555)$ is
$J/\psi \gamma$.
This decay has the advantage that
it could be isolated from the normal hadronic background
using the leptonic decay mode of the $J/\psi$.
\par
In this paper a search is presented for the \chic states in the reaction
\begin{equation}
pp \rightarrow p_{f} (J/\psi \gamma) p_{s}
\label{eq:b}
\end{equation}
with $J/\psi$~$\rightarrow$~$e^+e^-$.
Reaction~(\ref{eq:b})
has been isolated
from the sample of events having four
outgoing
charged tracks and one isolated $\gamma$, not associated
with a charged track impact, reconstructed in the GAMS-4000
calorimeter,
by first imposing the following cuts on the components of
missing momentum:
$|$missing~$P_{x}| <  14.0$ GeV/c,
$|$missing~$P_{y}| <  0.20$ GeV/c and
$|$missing~$P_{z}| <  0.16$ GeV/c,
where the $x$ axis is along the beam
direction.
A correlation between
pulse-height and momentum
obtained from a system of
scintillation counters was used to ensure that the slow
particle was a proton.
\par
One or both of the centrally produced charged tracks are required to impact on
the calorimeter.
The shower profile associated with the
charged track is required to be consistent with
being an electromagnetic shower.
Fig.~\ref{fi:1} shows a plot of the energy deposited in the GAMS calorimeter
divided
by the momentum of the charged track detected in Omega. A clear peak can be
observed
centred at $E/p$~=~1.0 due to electrons. The electrons have
been selected by requiring
0.9~$\le$~$E/p$~$\le$1.1.
At least one charged track per event
is required to be identified as an electron.
If the other charged track hits the calorimeter it is required to be compatible
with being an electron ($E/p$~$\ge$~0.8).
Fig.~\ref{fi:2}a) shows the resulting $e^+e^-$ mass spectrum which peaks
near zero consistent with the the majority of the electrons being due to
$\gamma$ conversions.
These $\gamma$ conversions were selected by requiring
M($e^+e^-)$~$\le$~0.1~GeV.
The momentum vector of the converted $\gamma$ has been combined with the
$\gamma$ reconstructed in GAMS. Fig.~\ref{fi:2}b) shows the resulting $\gamma
\gamma$ mass
spectra where clear peaks can be observed at the $\pi^0$ and $\eta$ masses.
\par
Fig.~\ref{fi:3}a) shows the
$e^+e^-$ mass spectrum for m($e^+e^-)$~$\ge$~2.0 GeV.
There is no evidence for a statistically significant peak
at the mass of the $J/\psi$.
Superimposed on the mass spectrum is a Monte Carlo prediction for a $J/\psi$
peak
coming from \chic decays.
Possible $J/\psi$ events have been selected by requiring
3.05~$\le$~M($e^+e^-)$~$\le$~3.15~GeV.
Fig.~\ref{fi:3}b) shows the resulting $J/\psi \gamma$ mass
spectra.
The mass resolution
has been calculated from Monte Carlo to be $\sigma$~=~55 MeV.
\par
Since
there is no significant
evidence for \chic production
only an upper limit
can be calculated.
After correcting for
geometrical acceptances, detector efficiencies,
losses due to cuts,
and unseen $J/\psi$ decay modes,
the cross-section for
the \chic resonances decaying to $J/\psi \gamma$
at $\sqrt s$~=~29.1~GeV is
$\sigma$(\chic$ \rightarrow J/\psi \gamma$)~$<$~ 2.0~nb (90 \% CL).
\par
A search has also been made for the reaction
\begin{equation}
pp \rightarrow p_{f} (J/\psi ) p_{s}
\label{eq:c}
\end{equation}
with $J/\psi$~$\rightarrow$~$e^+e^-$.
Reaction~(\ref{eq:c})
has been isolated
from the sample of events having four
outgoing
charged tracks
by first imposing the cuts on the components of
missing momentum described above.
Fig.~\ref{fi:4}a) shows a plot of the energy deposited in the GAMS calorimeter
divided
by the momentum of the charged track detected in Omega. In this case there
is a shoulder at 1.0
due to electrons. Electron candidates have
been selected by requiring
0.9~$\le$~$E/p$~$\le$1.1.
At least one charged track per event
is required to be identified as an electron.
If the other charged track hits the calorimeter it is required to be compatible
with being an electron ($E/p$~$\ge$~0.8).
Fig.~\ref{fi:4}b) shows the resulting $e^+e^-$ mass spectrum for
M($e^+e^-$) above 2~GeV. There is no sign of a peak in the
$J/\psi$ region.
\par
A study of other final states has been performed in order to search for
$J/\psi \pi^0$, $J/\psi \eta$, $J/\psi \pi^+\pi^-$ and $J/\psi
\pi^0\pi^0$.
In all cases there is no sign for a peak in the $ e^+e^-$ mass spectrum
at the $J/\psi$ mass.
\par
In order to search for the production of the $\eta_c(1S)$
we have studied the channels constituting its dominant decay modes, namely:
\etapipi and \kkpi.
The selection of the \etapipi and \kkpi channels have been described
in refs.~\cite{f1pap} and \cite{kkpipap} respectively.
Fig.~\ref{fi:4}c) shows the \etapipi mass spectrum and
fig.~\ref{fi:4}d) shows the \kkpi mass spectrum above 2~GeV.
There is no evidence for any $\eta_c(1S)$ signal above the background.
\par
In summary,
a search for centrally produced charmonium states has been presented.
There
is no significant
evidence for
any charmonium production.
In particular we have calculated an upper limit of 2~nb for
the cross section for \chic production using
the decay $\chi_c(1P)\rightarrow J/\psi\gamma$ with the
$J/\psi$ decaying to $e^+e^-$.
This upper limit could
be used as a test of the
hypothesis that
the soft Pomeron
has a very weak coupling to $c \overline c$ pairs~\cite{DL2}.
There is also evidence
that in central $pp$ collisions $s \bar{s}$ production is much weaker than
$n \bar{n}$ production. This evidence comes from the fact that
the cross section for the production of the $f_2(1270)$, whose production
has been found to be consistent with DPE~\cite{pipikkpap}, is
more than 40 times greater than the cross section of the $f_2^\prime(1525)$.
Hence there could be some strong dependence on
the mass of the quarks which could explain the lack of $c \bar{c}$ in DPE.
\begin{center}
{\bf Acknowledgements}
\end{center}
\par
This work is supported, in part, by grants from
the British Particle Physics and Astronomy Research Council,
the British Royal Society,
the Ministry of Education, Science, Sports and Culture of Japan
(grants no. 07044098 and 1004100), the French Programme International
de Cooperation Scientifique (grant no. 576)
and
the Russian Foundation for Basic Research
(grants 96-15-96633 and 98-02-22032).

\clearpage
{ \large \bf Figures \rm}
\begin{figure}[h]
\caption{The energy associated with a charged track impact
deposited in the GAMS-4000 calorimeter divided by the momentum of the
track measured in Omega for reaction~(\ref{eq:b}).
}
\label{fi:1}
\end{figure}
\begin{figure}[h]
\caption{a) The $e^+e^-$ mass spectrum.
b) The $\gamma \gamma$ mass spectrum when one of the
$\gamma$ is observed decaying to $e^+e^-$.
}
\label{fi:2}
\end{figure}
\begin{figure}[h]
\caption{a) The $e^+e^-$ mass spectrum.
Superimposed is the $J/\psi$ signal expected from a Monte Carlo simulation.
b) The $e^+e^-  \gamma$ mass spectrum for
3.05~$\le$~M($e^+e^-)$~$\le$~3.15~GeV.
%The curve is the result of the fit described in the text.
}
\label{fi:3}
\end{figure}
\begin{figure}[h]
\caption{The energy associated with a charged track impact
deposited in the GAMS-4000 calorimeter divided by the momentum of the
track measured in Omega for reaction~(\ref{eq:c}) and
b) the $e^+e^-$ mass spectrum.
c) The \etapipi and d) \kkpi mass spectra.
}
\label{fi:4}
\end{figure}
\begin{center}
\epsfig{figure=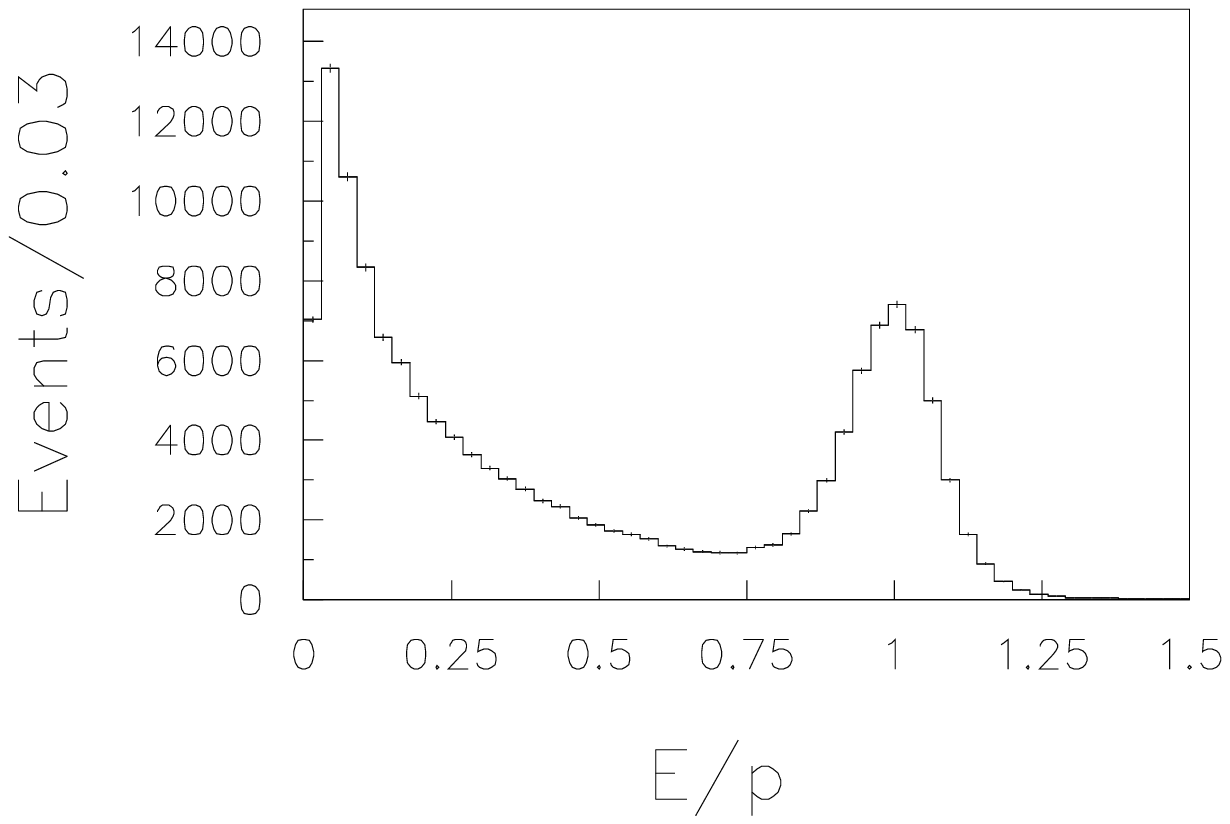,height=22cm,width=17cm}
\end{center}
\begin{center} {Figure 1} \end{center}
\newpage
\begin{center}
\epsfig{figure=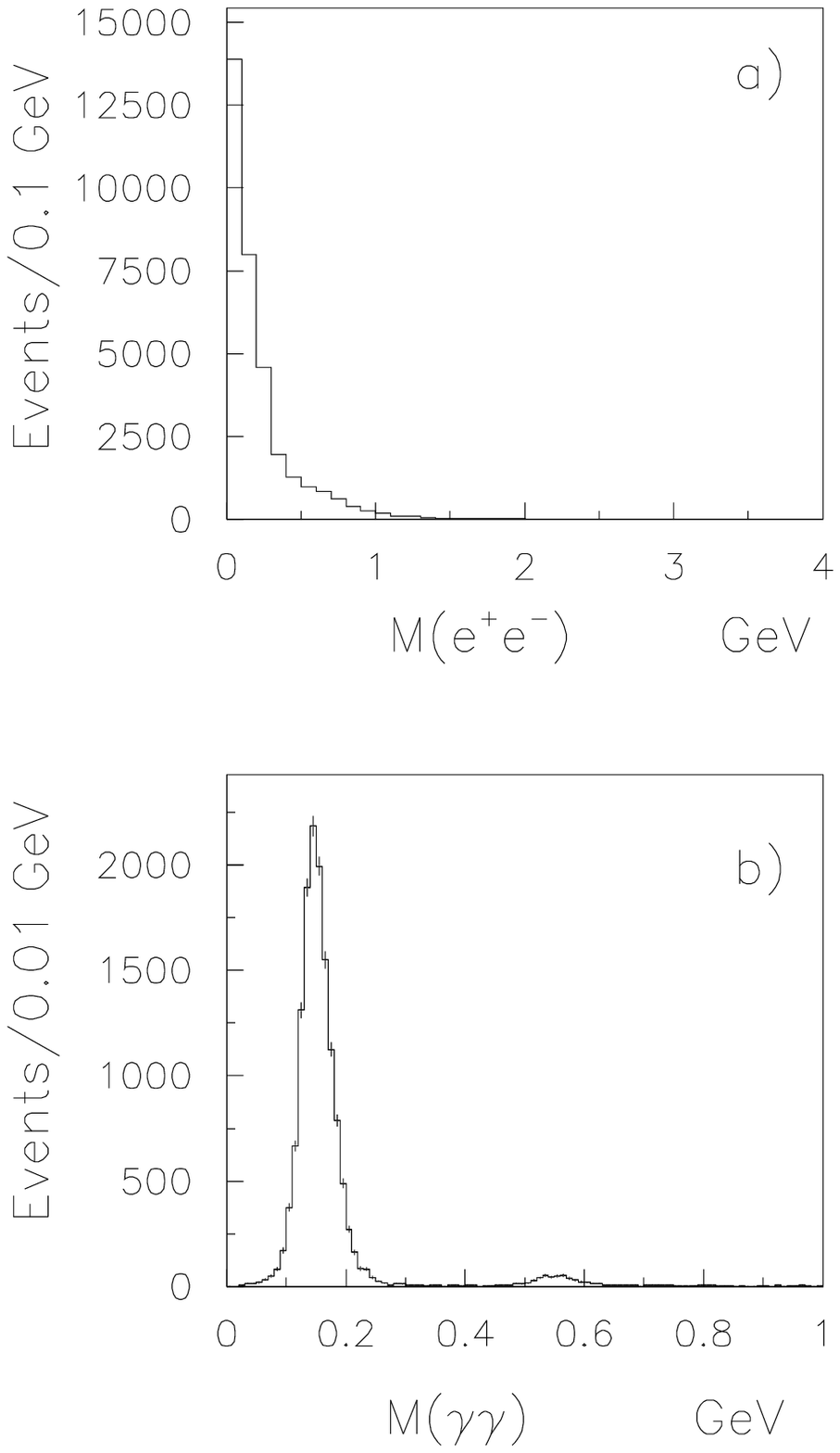,height=22cm,width=17cm}
\end{center}
\begin{center} {Figure 2} \end{center}
\newpage
\begin{center}
\epsfig{figure=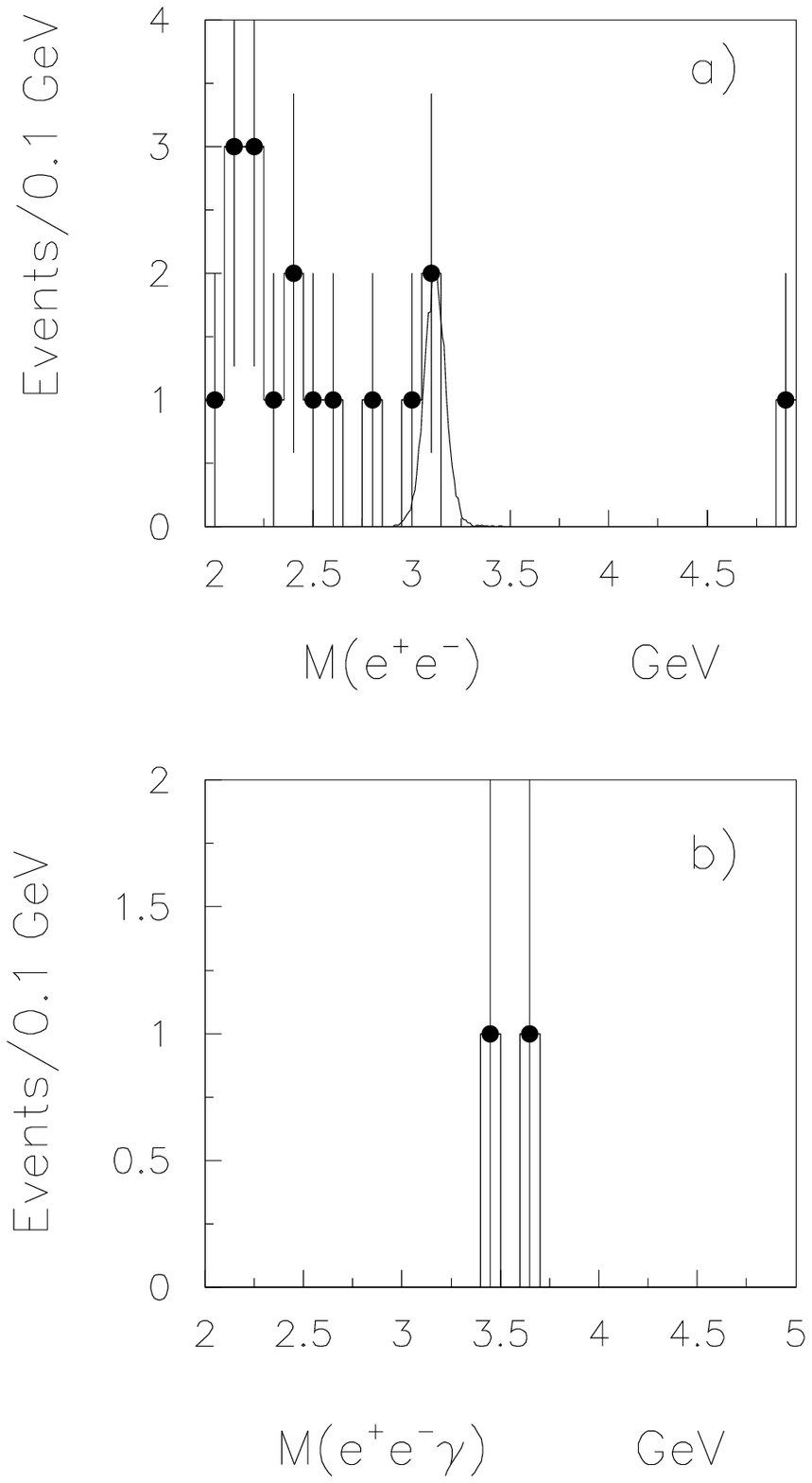,height=22cm,width=17cm}
\end{center}
\begin{center} {Figure 3} \end{center}
\newpage
\begin{center}
\epsfig{figure=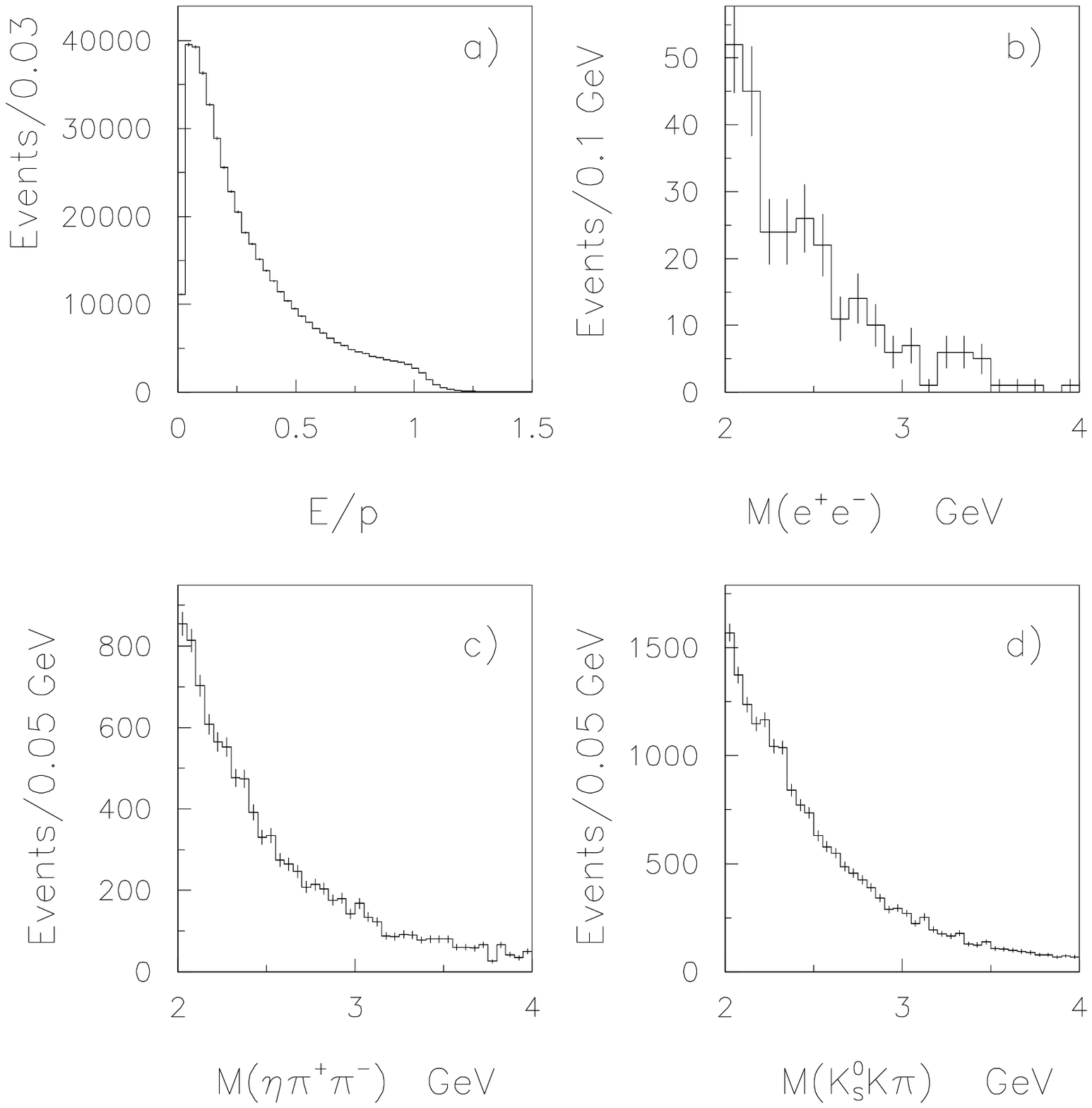,height=22cm,width=17cm}
\end{center}
\begin{center} {Figure 4} \end{center}
\end{document}